\documentclass[aps,pra,twocolumn,superscriptaddress,showpacs,floatfix]{revtex4-1}

\usepackage{graphicx}
\usepackage{amsmath}
\usepackage{cleveref}
\usepackage{color}

\usepackage[bookmarks=true,colorlinks=true,urlcolor=blue,linkcolor=blue,citecolor=blue,breaklinks]{hyperref}

\usepackage{amssymb}
\usepackage{booktabs}

\begin{document}


\title{Electron-mediated phonon-phonon coupling drives the vibrational relaxation of CO on Cu(100)}


\newcommand*{\fnref}[1]{\textsuperscript{\ref{#1}}}

\newcommand\myeq{\mathrel{\overset{\makebox[0pt]{\mbox{\normalfont\tiny\sffamily $\Gamma\rightarrow0$}}}{=}}}

\newcommand*{\DIPC}[0]{{
Donostia International Physics Center (DIPC),
Paseo Manuel de Lardizabal 4, 20018 Donostia-San Sebasti\'an, Spain}}

\newcommand*{\FUBER}{{
Institut f{\"u}r Chemie und Biochemie, Freie Universit{\"a}t Berlin,
Takustr.~3, 14195 Berlin, Germany}}

\newcommand{\CFM}[0]{{
Centro de F\'{\i}sica de Materiales CFM/MPC (CSIC-UPV/EHU), Paseo Manuel de Lardizabal 5, 20018 Donostia-San Sebasti\'an, Spain}}

\newcommand{\QUIMICAS}[0]{{
Departamento de F\'{\i}sica de Materiales, Facultad de Qu\'{\i}micas UPV/EHU,
Apartado 1072, 20080 Donostia-San Sebasti\'an, Spain}}

\author{D. Novko}
\email{dino.novko@gmail.com}
\affiliation{\FUBER}
\affiliation{\DIPC}
\author{M. Alducin}
\affiliation{\CFM}
\affiliation{\DIPC}
\author{J.I. Juaristi}
\affiliation{\QUIMICAS}
\affiliation{\CFM}
\affiliation{\DIPC}


\date{\today}

\begin{abstract}

We bring forth a consistent theory for the electron-mediated vibrational intermode coupling that clarifies the microscopic mechanism behind the vibrational relaxation of adsorbates on metal surfaces. Our analysis points out the inability of state-of-the-art nonadiabatic theories to quantitatively reproduce the experimental linewidth of the CO internal stretch mode on Cu(100) and it emphasizes the crucial role of the electron-mediated phonon-phonon coupling in this regard. The results demonstrate a strong electron-mediated coupling between the internal stretch and low-energy CO modes, but also a significant role of surface motion. Our nonadiabatic theory is also able to explain the temperature dependence of the internal stretch phonon linewidth, thus far considered a sign of the direct anharmonic coupling.

\end{abstract}


\maketitle




\textit{Introduction.--} Most of the theories describing dynamical processes at surfaces rely on the validity of the adiabatic Born-Oppenheimer approximation. However, in the case of metal surfaces, high concentrations of conducting electrons can, in principle, exchange energy with the adsorbate nuclear degrees of freedom. In fact, there is growing experimental evidence that points out the existence of such nonadiabatic effects~\cite{bib:rettner85,bib:ryberg85,bib:persson85,bib:morin92,bib:germer94,bib:ueba97,bib:cooper10,bib:shirhatti14}. Among them, the significant vibrational linewidths of molecules adsorbed on metal surfaces, which are reported in infrared absorption or pump-probe spectroscopy experiments, are considered to be clear fingerprints of the electron-mediated vibrational relaxation~\cite{bib:ryberg85,bib:persson85,bib:morin92,bib:germer94,bib:ueba97}. Besides, the nonadiabatic coupling underlies a variety of surface reactions and adsorbate motions (e.g., vibrations, rotations, and lateral hopping) induced by means of inelastic electron currents~\cite{bib:stipe97,bib:stipe98,bib:komeda02} and femtosecond laser pulses~\cite{bib:bonn00,bib:backus05,bib:inoue16}.

Apart from nonadiabatic coupling, a key ingredient for depicting the aforesaid processes is the intermode vibrational coupling~\cite{bib:persson85,bib:stipe98,bib:komeda02,bib:inoue16}. For example, hot-electron induced nonequilibrium dynamics of CO on Cu(100), which is studied by time-resolved vibrational sum-frequency generation, reveals the importance of intermode coupling in the CO desorption process~\cite{bib:inoue16}. Furthermore, the temperature dependence of the CO internal stretch (IS) mode linewidth studied by infrared spectroscopy was explained in terms of the coupling with the thermally-excited low-energy (LE) CO vibrational modes~\cite{bib:persson85,bib:germer94}.

The ensuing theoretical efforts aimed to comprehend these experimental observations are remarkable. In particular, the nonadiabatic relaxation of vibrationally excited adsorbates  is most commonly studied either by relaxation-rate calculations based on first-order perturbation theory~\cite{bib:persson80,bib:hellsing84,bib:headgordon92,bib:askerka16,bib:rittmeyer17} or by performing molecular dynamics with the corresponding electronic friction~\cite{bib:saalfrank14,bib:rittmeyer15}. In the former case, the intermode coupling is usually tackled by including an additional damping rate due to direct anharmonic coupling~\cite{bib:persson84,bib:persson02}. Despite the valuable qualitative insight gained, these theories are still unable to give precise quantitative estimations of the experimental vibrational relaxation rates and, hence, neither can they clarify which relevant mechanisms are ruling them. An emblematic example is the IS mode of CO on Cu(100). Even if its relaxation is considered to be mostly nonadiabatic~\cite{bib:ryberg85,bib:saalfrank06}, the corresponding vibrational linewidths calculated recently with \emph{ab initio} first-order perturbation theory are significantly lower than the experimental values~\cite{bib:maurer16,bib:novko16a}. This discrepancy together with the indisputable evidence for the intermode coupling in CO/Cu(100)~\cite{bib:germer94, bib:inoue16} points out to an overlooked relaxation mechanism that should incorporate both the nonadiabatic and intermode couplings on the same footing.

\begin{figure}[t]
\includegraphics[width=0.485\textwidth]{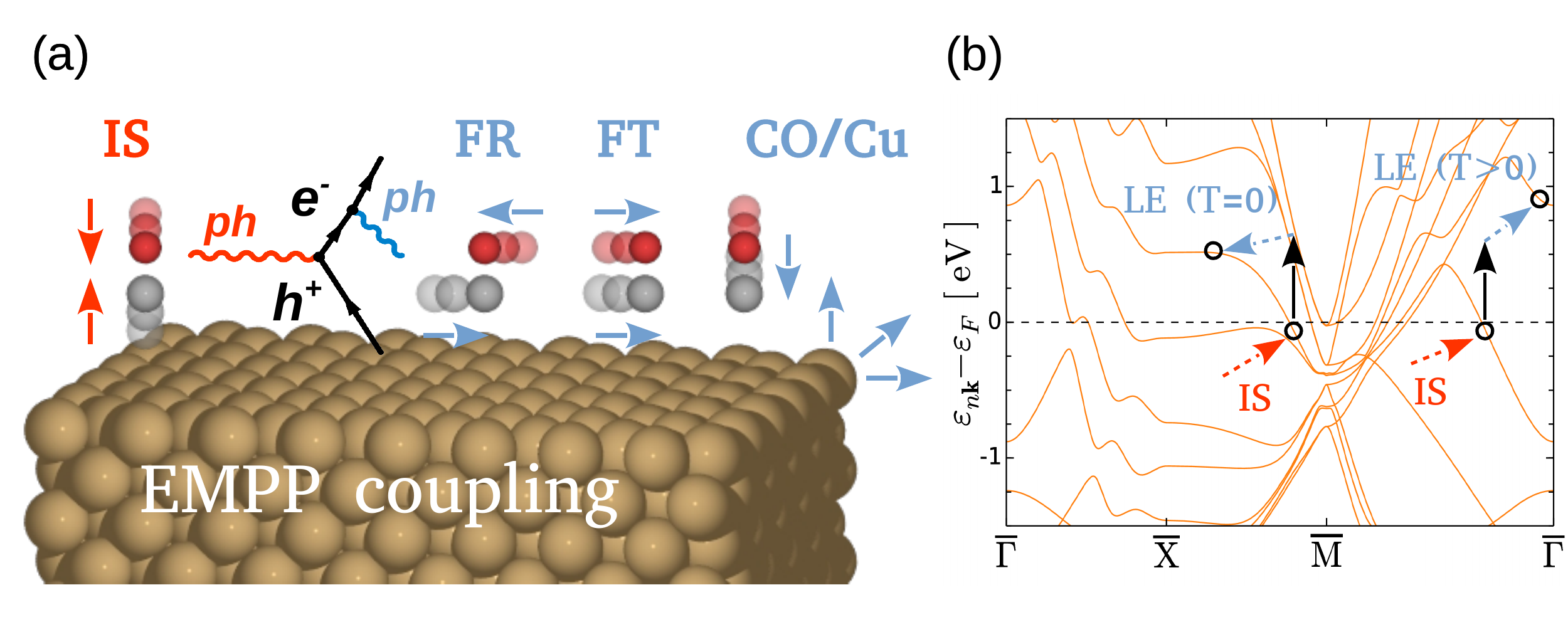}
\caption{\label{fig:fig1}(a) Schematic of the EMPP coupling process of the IS mode of CO on Cu(100). The IS mode (red) couples to low-energy FR and FT modes, as well as to the modes consisting of the joint motion of CO and Cu surface atoms (blue). (b) Indirect phonon-assisted electronic transitions involved in the EMPP process at $T=0$\,K (phonon emission) and $T>0$\,K (phonon absorption). In each case, both intraband and interband scattering are allowed.}
\end{figure}

In this Letter we demonstrate how nonadiabatic coupling can naturally account for intermode coupling if the former is treated up to second-order in the electron-phonon interaction. This theory that includes the process conjoining nonadiabaticity and intermode transitions --known as electron-mediated phonon-phonon (EMPP) coupling-- can  correctly describe the mechanisms behind the vibrational relaxation of ordered molecules on metal surfaces. Using the prototypical IS mode of CO on Cu(100), we show that the present  phonon linewidth formula combined with first-principles methodologies is finally able to explain the experimental relaxation rates. Importantly, our results show that the EMPP process dominates over the commonly-used first-order nonadiabatic contribution. Specifically, IS couples strongly via electron-hole (\emph{e-h}) pairs to other LE molecular phonon modes, i.e., to the frustrated rotation (FR) and frustrated translation (FT) modes [see Fig.\,\ref{fig:fig1}(a)]. Even more surprising, surface motion also plays an important role in the electron-mediated vibrational relaxation. Finally, as another remarkable success, our theory explains the temperature dependence of the IS mode linewidth. This result proves that the temperature dependence is not only triggered by the usual direct phonon-phonon coupling and that the commonly overlooked electron-mediated processes are also a significant factor in this respect.

\textit{State-of-the-art theory.--} Thus far the vibrational relaxation of molecules at metal surfaces was studied mostly by means of the (first-order) Fermi's golden rule formula (i.e., by only considering the terms $\propto g^2$, where $g$ is the electron-phonon matrix element)~\cite{bib:persson80,bib:hellsing84,bib:rantala86,bib:headgordon92,bib:krishna06,bib:forsblom07,bib:askerka16,bib:maurer16}. An analogous pathway in many-body perturbation theory (MBPT) is to calculate the phonon self-energy due to electron-phonon coupling up to first order, $\pi_{\lambda}^{[1]}(\mathbf{q},\omega)$ ($\lambda$, $\mathbf{q}$, and $\omega$ are the index, momentum, and energy of the phonon mode, respectively)~\cite{bib:novko16a}. The corresponding phonon linewidth is obtained by taking the imaginary part of $\pi_{\lambda}^{[1]}$, i.e., $\gamma_{\mathbf{q}\lambda}^{[1]}=-2\mathrm{Im}\,\pi_{\lambda}^{[1]}(\mathbf{q},\omega_{\mathbf{q}\lambda})$. In infrared spectroscopy, light is directly exciting only the long-wavelength phonons (i.e., $\mathbf{q}\approx 0$). Thus, here we focus on the $\mathbf{q}\approx 0$ part of $\pi_{\lambda}^{[1]}$, which is given by
\begin{eqnarray}
\pi_{\lambda}^{[1]}(\omega)=\sum_{\mu\mu'\mathbf{k}}\left| g_{\lambda}^{\mu\mu'}(\mathbf{k},0) \right|^2\frac{f(\varepsilon_{\mu\mathbf{k}})-f(\varepsilon_{\mu'\mathbf{k}})}{\omega+\varepsilon_{\mu\mathbf{k}}-\varepsilon_{\mu'\mathbf{k}}+i\eta},
\label{eq1}
\end{eqnarray}
where $\mu$, $\mathbf{k}$, and $\varepsilon_{\mu\mathbf{k}}$ are the electron band index, momentum, and energy, respectively. The Fermi-Dirac distribution function is defined as $f(\varepsilon_{\mu\mathbf{k}})=1/(e^{\beta(\varepsilon_{\mu\mathbf{k}}-\varepsilon_F)}+1)$, where $\beta=1/(k_BT)$, $k_B$ is the Boltzmann constant, $T$ is the temperature and $\varepsilon_F$ is the Fermi energy. In the limit $\eta\rightarrow 0^{+}$ one obtains the exact first-order phonon linewidth formula~\cite{bib:novko16a}, in which the intraband part ($\mu=\mu'$) of Eq.\,\eqref{eq1} vanishes and the only remaining contribution comes from the direct interband excitations ($\mu\neq\mu'$).

\begin{figure}[b]
\includegraphics[width=0.5\textwidth]{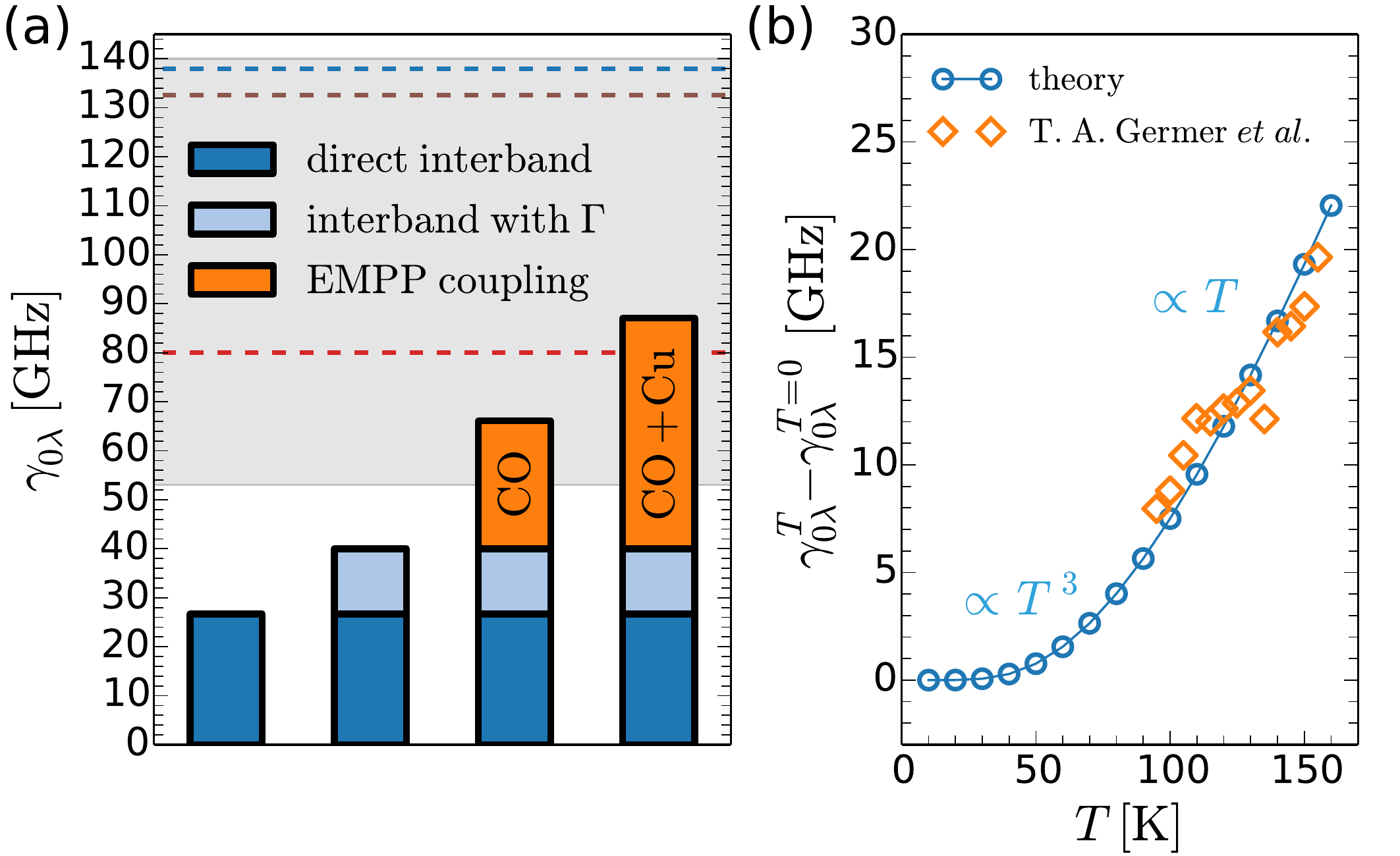}
\caption{\label{fig:fig2}(a) Different contributions to the IS phonon linewidth $\gamma_{0\lambda}$ coming from direct interband transitions (dark blue), interband transitions with phenomenological broadening $\Gamma=60$\,meV (light blue), and the EMPP coupling at $T=10$\,K (orange). The experimental total linewidths obtained in Refs.\,\cite{bib:ryberg85} and \cite{bib:morin92} are shown with blue and brown dashed lines, respectively. The inelastic contribution obtained in Ref.\,\cite{bib:morin92} is shown with a red dashed line. Gray shaded area represents all the other experimental values, including experimental error bars~\cite{bib:ueba97}. (b) Temperature dependence of the IS phonon linewidth as obtained from the EMPP coupling term (blue dots) and the corresponding experimental values~\cite{bib:germer94} (orange diamonds).} 
\end{figure}

Early parameter-dependent calculations of the IS linewidth based on the Fermi's golden rule reported good agreement with experiments~\cite{bib:persson80,bib:hellsing84,bib:rantala86,bib:headgordon92,bib:krishna06,bib:forsblom07}. However, here we show that Eq.\,\eqref{eq1} is not enough to explain the experimental linewidths~\footnote{The considered good quantitative agreement with experiments might be an artifact of the parameter-based electronic structure calculations, i.e., large values of broadening $\eta$, or different approximations used (quasi-elastic limit, etc.)~\cite{bib:novko16a}}. In fact, our \emph{ab initio} calculations of Eq.\,\eqref{eq1} for c$(2\times2)$-CO/Cu(100) based on accurate density functional perturbation theory~\cite{bib:baroni01,bib:qe} give $\gamma_{0\lambda}^{[1]}=26.7$\,GHz (see Ref.\,\cite{bib:si} for computational details), which is far from the experimental values $\gamma_{\mathrm{exp.}}>50$\,GHz~\cite{bib:ryberg85,bib:morin92} [see Fig.\,\ref{fig:fig2}(a)]. The discrepancy between the phonon linewidth obtained from Eq.\,\eqref{eq1} and in experiments has also been discussed recently~\cite{bib:askerka16,bib:maurer16,bib:novko16a}. Note in passing that our value $\gamma_{0\lambda}^{[1]}$ is in almost perfect agreement with the result obtained within a mixed quantum-classical theory when an infinite electron coherence time $\tau_e$ is used~\cite{bib:rittmeyer17}.

As shown in Ref.\,\cite{bib:novko16a}, the agreement with the experimental linewidth can be improved by taking into account interband electronic scattering processes (i.e., electron-impurity, electron-phonon, or electron-electron scattering). Phenomenologically, this can be done by replacing the infinitesimal $\eta$ in Eq.\,\eqref{eq1} with a finite broadening $\Gamma$. By using a physically motivated $\Gamma=60$\,meV~\cite{bib:brown16}, we get $\gamma_{0\lambda}^{[1]}=40$\,GHz [second column in Fig.\,\ref{fig:fig2}(a)]. However, to achieve agreement with experiments one needs to use non-physically large $\Gamma$ values, i.e., much larger than 100 meV. In a similar way, the agreement with experiments in Ref.\,\cite{bib:rittmeyer17} is improved by using a finite $\tau_e$ value (note that $\tau_e=2\pi\hbar/\Gamma$).

%
%
For completeness, we have additionally performed \emph{ab initio} molecular dynamics simulations with electronic friction (AIMDEF) as in Refs.\,\cite{bib:juaristi08,bib:blanco14,bib:saalfrank14,bib:novko15} to calculate the energy relaxation of the IS mode. Position-dependent electronic friction coefficients are calculated within the local density friction and independent atom approximations~\cite{bib:juaristi08} (see Ref.\,\cite{bib:si} for computational details). By restricting the CO dynamics along the surface normal on the rigid surface, a value $\gamma_{\mathrm{AIMDEF}}=18.4$\,GHz is obtained, which is in quite fair agreement with the result obtained from Eq.\,\eqref{eq1}, and, hence, it also fails to reproduce the experimental linewidths.

All the aforesaid results highlight that additional channels must be behind the large experimental IS relaxation rate. In particular, the non-negligible temperature dependence of the IS mode linewidth reported in infrared spectroscopy~\cite{bib:germer94} and the recent experiments on CO hot-electron induced nonequilibrium dynamics~\cite{bib:inoue16} suggest that a key role is played by intermode transitions, for which a more rigorous study is required.

\textit{Electron-mediated phonon-phonon coupling.--} The usual procedure for treating intermode transitions is to consider the direct anharmonic coupling between two phonon modes~\cite{bib:persson84,bib:persson02}. However, for the high-energy (HE) IS mode ($\omega_{\mathrm{exp.}}=62.54\,\mathrm{THz}=256\,\mathrm{meV}$~\cite{bib:ryberg85}), direct anharmonic coupling with the LE phonon modes, such as FR ($\omega_{\mathrm{exp.}}=8.54\,\mathrm{THz}=35.32\,\mathrm{meV}$~\cite{bib:hirschmugl90}), is very inefficient due to the large difference in energy~\cite{bib:persson84}~\footnote{We also checked the minor role of the direct anharmonic coupling by performing six-dimensional AIMD simulations in which the CO is initially excited in the IS mode. A similar conclusion is extracted from the Supplemental Material of Ref.\,\cite{bib:rittmeyer17}.}. 
Nevertheless, the presence of the \emph{e-h} pair continuum coming from the metal surface can compensate this energy mismatch and allow for the EMPP coupling. In other words, the HE mode can excite \emph{e-h} pairs which in turn can undergo further electron-phonon scattering with the LE modes [see Fig.\,\ref{fig:fig1}(a)]. In the context of MBPT, the EMPP process is accounted for by calculating the second-order intraband phonon self-energy due to the electron-phonon interaction~\cite{bib:si,bib:marsiglio92,bib:maksimov96,bib:cappelluti06},
\begin{eqnarray}
&&\pi_{\lambda}^{[2]}(\omega)=-\sum_{\mu\mu'\mathbf{k}\lambda'\mathbf{k}'}
\left| g_{\lambda}^{\mu\mu}(\mathbf{k},0)\right|^2\left[1-\frac{g_{\lambda}^{\mu'\mu'}(\mathbf{k}',0)}{g_{\lambda}^{\mu\mu}(\mathbf{k},0)}\right]\nonumber\\
&&\times\left| g_{\lambda'}^{\mu\mu'}(\mathbf{k},\mathbf{q}')\right|^2\sum_{s,s'=\pm 1}\frac{f(\varepsilon_{\mu\mathbf{k}})-f(\varepsilon_{\mu'\mathbf{k}'}-s's\omega_{\mathbf{q}'\lambda'})}{\varepsilon_{\mu\mathbf{k}}-(\varepsilon_{\mu'\mathbf{k}'}-s's\omega_{\mathbf{q}'\lambda'})}\nonumber\\
&&\times\frac{s\left[ n_b(s\omega_{\mathbf{q}'\lambda'})+f(s'\varepsilon_{\mu'\mathbf{k}'}) \right]}{\omega\left[ \omega+i\eta+s'(\varepsilon_{\mu\mathbf{k}}-\varepsilon_{\mu'\mathbf{k}'})+s\omega_{\mathbf{q}'\lambda'} \right]},
\label{eq2}
\end{eqnarray}
where $n_b(\omega_{\mathbf{q}\lambda})=1/(e^{\beta\omega_{\mathbf{q}\lambda}}-1)$ is the Bose-Einstein distribution function~\footnote{Note that $n_b(-\omega_{\mathbf{q}\lambda})=-1-n_b(\omega_{\mathbf{q}\lambda})$ and $f(-\varepsilon_{\mu\mathbf{k}})=1-f(\varepsilon_{\mu\mathbf{k}})$} and $\mathbf{q'=k'-k}$. As noted before, in infrared spectroscopy the IS mode is not able to excite direct intraband transitions due to the vanishingly small phonon momentum. Thus, the first non-vanishing intraband contribution is Eq.\,\eqref{eq2}, which breaks down the momentum conservation by coupling the studied $(\mathbf{q}\approx 0,\lambda)$ mode with other $(\mathbf{q}',\lambda')$ phonon modes via \emph{e-h} pairs [see Fig.\,\ref{fig:fig1}(b)]. Such indirect phonon-assisted electronic transitions~\cite{bib:kupcic09,bib:giustino17,bib:novko17} involved in this process are considered to be an important mechanism in hot-carrier generation via plasmon decay in Cu and other noble metals~\cite{bib:brown16}.

\begin{table*}[t]
\caption{\label{tab:table1}Energy range (THz) of the c$(2\times2)$-CO/Cu(100) phonon modes in the 1st Brillouin zone ($\omega_{\mathbf{q}'\lambda'}$), phonon energies for $\mathbf{q}\approx 0$ ($\omega_{0\lambda'}$), and the corresponding experimental values. The last row shows contributions to the EMPP coupling term of the IS phonon linewidth $\gamma_{0\lambda}^{[2]}$ (GHz) coming from different $\lambda'$ modes at $T=10$\,K ($T=160$\,K).}
\begin{ruledtabular}
\begin{tabular}{cccccc}
 &\multicolumn{5}{c}{$\lambda'$ mode ($\lambda=$ IS)}\\
 & IS & ES & FR & FT & CO/Cu \\ \hline
\rule{0pt}{3ex}
$\omega_{\mathbf{q}'\lambda'}$ $\mathrm{[THz]}$ & 58.70 - 61.37 &  12.11 - 12.33  & 9.14 - 9.78 & 5.79 - 6.33 & $< 7$ \\
$\omega_{0\lambda'}$ $\mathrm{[THz]}$ & 61.37 &  12.33  & 9.78 & 6.05 & $< 7$ \\
Exp. $\omega_{0\lambda'}$ $\mathrm{[THz]}$ &  62.54 \cite{bib:ryberg85}  & 10.34 \cite{bib:hirschmugl90}  & 8.54 \cite{bib:hirschmugl90} & 0.96 \cite{bib:graham03}  & $\lesssim 7$ \cite{bib:nilsson73} \\
$\gamma_{0\lambda}^{[2]}$ $\mathrm{[GHz]}$ & 0.48 (1.04) & 1.46 (1.56) & 15.58 (17.73) & 8.61 (12.60) & 20.96 (36.20) \\
\end{tabular}
\end{ruledtabular}
\end{table*}

With very few exceptions these kinds of electron-mediated, second-order processes have not been considered in the context of vibrating molecules at metal surfaces. In these few references, effects due to \emph{e-h} pair dephasing~\cite{bib:morawitz87} and direct anharmonic coupling followed by \emph{e-h} pair excitation and vice versa~\cite{bib:persson02,bib:ueba08b,bib:ueba08a} were investigated with parametrized models that only accounted for quasielastic electron scattering and that were fitted to match the experimental data. In contrast to these studies, here we include and calculate from first principles both the elastic and inelastic (i.e., fully dynamical) electron-phonon scattering~\cite{bib:si}.

Table\,\ref{tab:table1} (last row) shows the results for the EMPP coupling term of the IS phonon linewidth [$\gamma_{0\lambda}^{[2]}=-2\mathrm{Im}\,\pi_{\lambda}^{[2]}(\omega_{0\lambda})$] decomposed into contributions coming from different CO/Cu(100) phonon modes: IS ($\mathbf{q}'>0$)~\footnote{This ``self'' contribution to $\gamma_{0\lambda}^{[2]}$ comes from all the IS modes with $\mathbf{q}'$ in the 1st Brillouin zone that satisfy the condition $\omega_{\mathbf{q}'\lambda}<\omega_{0\lambda}$. In other words, this accounts for the coupling of the studied IS mode at $\mathbf{q}\approx0$ with all other $\mathbf{q}'$ modes within the same phonon band.}, CO-Cu or external stretch (ES), FR, FT, as well as modes consisting of the joint motion of CO and Cu(100) atoms (CO/Cu). We also provide the calculated phonon mode energies and compare them to available experimental data~\cite{bib:ryberg85,bib:hirschmugl90,bib:graham03,bib:nilsson73}. As Table\,\ref{tab:table1} shows, a significant damping of the IS mode is caused by the electron-mediated coupling with the FR modes within the molecular overlayer. On the other hand, the smallest contribution comes from the ES and the IS ($\mathbf{q}'>0$) modes. These results are actually in line with existing femtosecond laser experiments showing the importance of the IS-FR mode coupling in the desorption of CO from Cu(100)~\cite{bib:inoue16} and Ru(001)~\cite{bib:bonn00}. Similarly, the coupling between the IS and LE modes are thought to promote the surface hopping of CO on Pd(110) induced by inelastic tunneling electrons~\cite{bib:komeda02} and of CO on Pt(533) induced by a laser pulse~\cite{bib:backus05}. Hence, also in these cases our theory of EMPP coupling can be useful in elucidating what modes are actually involved in these kinds of surface reactions. Another important conclusion extracted from Table\,\ref{tab:table1} is that a considerable contribution to the IS phonon linewidth comes from the CO/Cu modes. This result contradicts the usual assumption that considers the coupling between the adsorbate IS mode and the modes involving surface motion irrelevant for the IS relaxation process, due to the large energy mismatch~\cite{bib:saalfrank06}. Here we show that such coupling is indeed possible whenever the continuum of surface conducting electrons is present to compensate the energy gap.

Next, we summarize in Fig.\,\ref{fig:fig2}(a) all the contributions to the phonon linewidth and compare them to the experimental findings. Blue and brown dashed lines represent the total experimental phonon linewidths obtained in infrared absorption spectroscopy~\cite{bib:ryberg85} and infrared pump-probe spectroscopy~\cite{bib:morin92}, respectively. The latter technique makes it possible to extract the energy relaxation contribution (i.e., inelastic process, red dashed line) from the total linewidth. The remaining contributions to the total phonon linewidth are elastic processes (i.e., dephasing via \emph{e-h} pairs~\cite{bib:morawitz87} and anharmonic coupling~\cite{bib:persson84}) and inhomogeneities (e.g., impurities and disorders). The gray shaded area covers the large dispersion of experimental data existing in the literature together with the experimental error bars~\cite{bib:ueba97}. In our theoretical framework, the interband contribution [Eq.\,\eqref{eq1}] is a purely inelastic process, while the EMPP coupling term [Eq.\,\eqref{eq2}] contains both inelastic (i.e., $\varepsilon_{\mu\mathbf{k}}\neq\varepsilon_{\mu'\mathbf{k}'}$) and elastic (i.e., $\varepsilon_{\mu\mathbf{k}}\approx\varepsilon_{\mu'\mathbf{k}'}$ as in \emph{e-h} pair dephasing~\cite{bib:morawitz87}) processes and both are together included in our calculations. 
All in all, the EMPP coupling term greatly improves upon the state-of-the-art theory, and is essential for understanding the experimental linewidths. Nevertheless, further work is desirable in order to reach the ultimate accuracy in the calculation of the phonon linewidth. Specifically, inclusion of inhomogeneities (e.g., electron-impurity scattering)~\cite{bib:novko16a} and electron correlation effects~\cite{bib:dou17} might improve the overall result.

\textit{Temperature dependence.--} 
The analysis of temperature effects in the CO IS mode also confirms the relevance of the EMPP coupling mechanism. Temperature  enters the usual (first-order) electron-phonon term [Eq.\,\eqref{fig:fig1}] only through the Fermi-Dirac distribution functions. In other words, the temperature is skewing the electron distribution and thus changing the rate of upward and downward electron transitions. However, the overall sum of these transitions [i.e., $f(\varepsilon_{\mu\mathbf{k}})-f(\varepsilon_{\mu'\mathbf{k}})$] is practically $T$-independent at moderate temperatures ($k_BT\ll\omega_{0\lambda}$)~\cite{bib:brandbyge95}. In our case, the linewidth of the CO IS mode coming from this term varies less than 1\,GHz for a temperature range of 40--300\,K~\cite{bib:novko16a}, which is also in accordance with recent calculations~\cite{bib:maurer16}. In contrast, temperature effects coming from the EMPP coupling [Eq.\,\eqref{eq2}], which enter the phonon linewidth through $n_b(\omega_{\mathbf{q}'\lambda'})$, are significantly more pronounced, as also suggested in studies of optical phonon modes in metals~\cite{bib:maksimov96,bib:cappelluti06}. Figure\,\ref{fig:fig2}(b) shows the results for the $T$-dependence of the IS phonon linewidth coming from the EMPP coupling [the zero-temperature value is $\gamma_{0\lambda}^{T=0}=87.1$\,GHz as in Fig.\,\ref{fig:fig2}(a)]. We observe that $\gamma_{0\lambda}^{[2]}$ increases as $T^3$ at low temperatures and linearly with $T$ at higher temperatures. The overall increase in the range $T=0-160$\,K is around 22\,GHz. Such a result remarks that these low  temperatures are enough to significantly enhance the population of the excited LE phonons and to thus cause an increase in the probability of electron-LE phonon scattering events. Table\,\ref{tab:table1} shows that the main contributors to the increase of the linewidth with temperature are the lowest-energy modes, i.e., CO/Cu and FT (see Ref.\,\cite{bib:si} for more details).

In Ref.\,\cite{bib:germer94} the $T$-dependence of the IS phonon linewidth is measured between 100--160\,K using infrared absorption spectroscopy (orange diamonds). The linear increase we obtain in that temperature range is in both qualitative and quantitative agreement with their findings~\cite{bib:si}. Usually, the $T$-dependence of the adsorbate HE modes is considered to be the footprint of an underlying pure dephasing mechanism, either coming from elastic scattering with LE modes (direct elastic anharmonic effect)~\cite{bib:persson84} or with \emph{e-h} pairs~\cite{bib:morawitz87}. 
The parametrized model of Ref.~\cite{bib:germer94} assigns the $T$-dependence to the former process and, then, the model parameters are adjusted to fit the experimental data. 
However, our results clearly show that it is the EMPP coupling the mechanism governing the $T$-dependence. Note that apart from inelastic nonadiabatic effects, this mechanism also includes the aforementioned \emph{e-h} pair dephasing when the studied $(\mathbf{q}\approx0,\lambda)$ mode is coupled to the different $\mathbf{q}'$ modes within the same $\lambda$ phonon band.


As a final remark, note that the dynamical models aimed to understand the experiments on CO desorption and surface hopping very often require including a phenomenological $T$-dependent relaxation-rate term (i.e., friction coefficient)~\cite{bib:brandbyge95,bib:inoue16}, even if the microscopic justification for such $T$-dependent friction is not clear enough. In searching a plausible explanation, Ueba and Persson constructed a parameter-based model in which an effective $T$-dependent friction, consisting of $T$-independent nonadiabatic coupling followed by direct anharmonic coupling, enters the heat transfer equation~\cite{bib:ueba08b,bib:ueba08a}. In this respect, our theory of EMPP coupling follows a similar line of reasoning. However, only this kind of theory that is entirely based on first principles and rigorous MBPT can finally provide new and detailed quantitative information on these $T$-dependent processes. In fact, we show that nonadiabaticity and anharmonic coupling are inseparable parts of the very same process if the electron-phonon coupling is treated consistently up to second order.

\textit{Conclusion.--} By combining many-body perturbation theory with density functional theory, we have investigated a new relaxation mechanism that bridges the nonadiabatic and vibrational intermode couplings, namely, the electron-mediated phonon-phonon coupling, which is then successfully applied to the prototypical case of the CO internal stretch mode on Cu(100). We have shown that the state-of-the-art nonadiabatic theory is not able to explain the internal stretch mode phonon linewidth obtained in infrared absorption spectroscopy and that inclusion of the electron-mediated phonon-phonon coupling is essential in this regard. The latter mechanism reveals a significant role of the low-energy modes in the relaxation process, such as CO frustrated rotation, but also the modes consisting of the joint motion of CO and surface atoms. Importantly, this new nonadiabatic mechanism is able to explain the temperature dependence of the internal stretch mode linewidth, which was hitherto ascribed to the direct anharmonic coupling or anharmonic dephasing processes. The proposed mechanism is quite general and can be used to elucidate which modes are involved in those surface reactions in which nonadiabatic effects and intermode coupling are expected to be relevant.

\begin{acknowledgments}
D.N. is grateful to I. Kup\v{c}i\'{c} for useful discussions and comments. Financial support by Donostia International Physics Center (DIPC) during various stages of this work is highly acknowledged.  M.\ A.\ and J.\ I.\ J.\ acknowledge the Spanish Ministerio de Econom\'{\i}a, Industria y Competitividad Grant No. FIS2016-76471-P. Computational resources were provided by the DIPC computing center.
\end{acknowledgments}

\newpage

\bibliography{emppc}

\end{document}